\documentclass[%
 reprint,
 amsmath,amssymb,
 aps,
]{revtex4-2}

\usepackage{graphicx}
\usepackage{dcolumn}
\usepackage{bm}

\begin{document}

\preprint{APS/123-QED}

\title{Switching waves-induced broadband Kerr frequency comb in fiber Fabry-Perot resonators}

\author{Thomas Bunel}
 \email{thomas.bunel@univ-lille.fr}
\author{Matteo Conforti}%
\affiliation{%
University of Lille, CNRS, UMR 8523-PhLAM Physique des Lasers, Atomes et Molécules, F-59000, Lille, France
}%

\author{Julien Lumeau}
\author{Antonin Moreau}
\affiliation{ 
Aix Marseille University, CNRS, Centrale Marseille, Institut Fresnel, Marseille, France
}%

\author{Arnaud Mussot}
\affiliation{%
 University of Lille, CNRS, UMR 8523-PhLAM Physique des Lasers, Atomes et Molécules, F-59000, Lille, France
}%

\date{\today}

\begin{abstract}
We report the generation of broadband frequency combs in fiber Fabry-Perot resonators in the normal dispersion regime enabled by the excitation of switching waves. We theoretically characterise the process by means of a transverse linear stability analysis of the Lugiato-Lefever equation, enabling precise prediction of the switching waves' frequencies. Experimentally, we employed a pulsed-pump fiber Fabry-Perot resonator operating in the normal dispersion regime, integrated into an all-fiber experimental setup. The synchronisation mismatch and the influence of dispersion is thoroughly discussed, unveiling the potential to generate a frequency comb spanning over 15~THz bandwidth, specifically leveraging a flattened low dispersion cavity.
\end{abstract}

\maketitle


\section{Introduction}

Nonlinear Kerr resonators have significantly advanced the generation of broadband optical frequency combs (OFCs) \cite{kippenberg_dissipative_2018,pasquazi_micro-combs_2018,sun_applications_2023} thriving in various applications including spectroscopy \cite{suh_microresonator_2016}, optical coherent tomography \cite{marchand_soliton_2021}, low noise radio frequency (RF) generation \cite{huang_broadband_2016}, distance ranging \cite{riemensberger_massively_2020} and high-speed optical telecommunication \cite{fulop_high-order_2018}. In this context, microresonators received intense interest owing to their potential for seamless integration into chip-scale integrated photonic circuits, high-quality factors, as well as high nonlinear performances \cite{kippenberg_dissipative_2018,pasquazi_micro-combs_2018,sun_applications_2023,brasch_photonic_2016}. Recently, fiber Fabry-Perot (FFP) resonators \cite{xiao_near-zero-dispersion_2023,nie_synthesized_2022,jia_photonic_2020,obrzud_temporal_2017} emerged as a promising alternative combining advantages of microresonator and fiber ring cavities \cite{leo_temporal_2010,englebert_temporal_2021}. They namely exhibit a high-quality factor, compact design, OFC with line-to-line spacing in the GHz range, simplified light coupling via FC/PC connectors, and ease of implementation within fiber systems, which is still a limitation faced by microresonators. Most of Kerr comb generation are based on cavity solitons generation in anomalous group velocity dispersion (GVD) resonators, resulting in a stable soliton recirculating in the resonator and a broad homogeneous OFC at the cavity output \cite{herr_temporal_2014,leo_temporal_2010}. 
However, normal GVD resonators are increasingly utilized for OFCs generation, in particular because of their high conversion efficiency between the pump and the generated comb lines \cite{fulop_high-order_2018,xue_mode-locked_2015,xue_normal-dispersion_2016,ji_engineered_2023,liu_stimulated_2022}. As the upper branch of a normally dispersive cavity does not exhibit modulation instability, various excitation techniques had to be identified to trigger OFCs in this regime, notably through mode-crossing effect \cite{xue_mode-locked_2015,xue_normal-dispersion_2016}, Brillouin effect \cite{postdeadline}, dual-pumping \cite{lucas_dynamic_2023}, coupled-cavity \cite{ji_engineered_2023,xue_coupled_cavities_2015,xue_normal-dispersion_2016}, modulated pump \cite{liu_stimulated_2022} or pulsed pumping scheme \cite{xiao_modeling_2023,li_experimental_2023,macnaughtan_temporal_2023,xu_frequency_2021,anderson_zero_2022}. 
All these tecnhiques rely on the generation of switching waves (SWs) \cite{xiao_modeling_2023,macnaughtan_temporal_2023,coen_convection_1999,parra-rivas_origin_2016,parra-rivas_coexistence_2017}, connecting the high and low stable states of bistable Kerr cavities. Switching waves may lock to each other, resulting in the generation of dark solitons  \cite{xue_mode-locked_2015,fulop_high-order_2018,nazemosadat_switching_2021,liang_generation_2014}. In the spectral domain, SWs manifest as a broadening characterized by two shoulders around the pump, resulting in a relatively flat OFC. While the study of SWs dates back to the 1980s \cite{rozanov_transverse_1982,rozanov_transverse_1982-1,coen_convection_1999},  there is a  renewed interest \cite{fulop_high-order_2018,liu_stimulated_2022,xiao_modeling_2023}, driven  by the recent technical advances in compact systems.


In this work, we report the generation of broadband Kerr frequency combs enabled by SWs in normal GVD FFP resonators through a pulsed pumping scheme. Leveraging an advanced all-fiber experimental setup, we can reliably produce highly stable SW-based OFCs. We delve into the dynamics of SWs generation, as a function of cavity detuning, synchronization mismatch, between the pulsed pump repetition rate and the cavity roundtrip time, and cavity dispersion. We present a theory to predict the SWs frequencies, which considers SWs as exact solutions of the Lugiato-Lefever equation (LLE) \cite{lugiato_spatial_1987}. This method is conceptually different from the previous studies, which interpret SWs spectral shoulders as  dispersive waves (DWs) emitted from a front \cite{macnaughtan_temporal_2023,xu_frequency_2021,ji_engineered_2023,xiao_modeling_2023,li_experimental_2023}. The experimental results shows very good agreement with numerical simulation and theoretical predictions.

\section{\label{theory}Theory}

\begin{figure}
\includegraphics{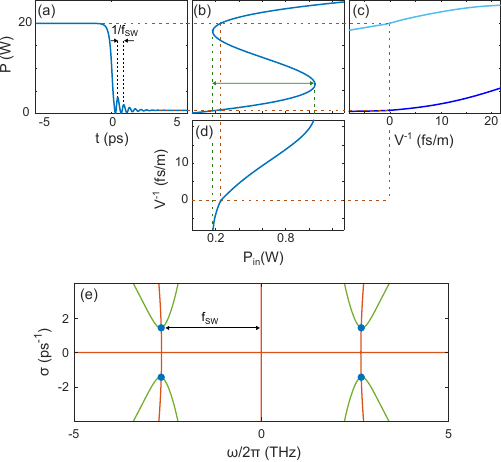}
\caption{\label{fig:SWtheo}Illustrations of switching waves dynamics using analytical predictions, for L$=16.18$~cm, $\mathcal{F}=600$, $\gamma=10.8~W^{-1}km^{-1}$, $\delta=0.032$~rad, and $P_{in}=0.5$~W. (a)~SW solution at its Maxwell point. The lower state exhibit oscillation with a frequency $f_{SW}$. (b)~Steady states of the system; the power range for which SWs are present is delimited by the green dotted lines and the green arrow. (c)~SWs inverse velocity as a function of lower and upper state power. (d)~SWs inverse velocity as a function of the input power. The orange dotted lines link the figures together showing that SWs generated at the MP have a velocity $V=0$. (e) Representation of the real part (green) and imaginary part (orange) of the Eq.~(\ref{disp}), with its roots (blue dots).}
\end{figure}

SWs are sharp fronts, propagating at a well-defined constant speed, which link the upper and lower states of a bistable regime. In this sense they are reminiscent of classical shock waves arising in fluid dynamics. The dispersion of the system tends to regularise sharp fronts with fast oscillation at specific frequencies, and in this sense they may be interpreted as dispersive shock waves \cite{Malaguti:14a}. Higher order dispersion (HOD) may induce SWs to shed resonant radiation (also called DW) \cite{Malaguti:14}. It is important to remark that SWs are exact solutions of LLE, and the presence of HOD is necessary to induce dispersive wave emission \cite{akhmediev_cherenkov_1995,wai_nonlinear_1986}. The oscillating front of a switching wave is part of the (nonlinear) solution itself, which exists even in absence of perturbations. However, several studies support the confusing idea that the oscillating fronts can be interpreted  as SW-enabled DWs \cite{macnaughtan_temporal_2023,xu_frequency_2021,ji_engineered_2023,xiao_modeling_2023,li_experimental_2023}. While this explanation has shown good alignment between experiments and theory, its description through DWs remains somewhat inadequate. 
Therefore, we propose an alternative approach by conducting a transverse stability analysis \cite{parra-rivas_coexistence_2017} to better describe this phenomenon. 

The starting point is the generalised Lugiato-Lefever equation \cite{lugiato_spatial_1987} 
in dimensional units:
\begin{align}
&i2L\frac{\partial A}{\partial z} + L  d(i\partial_t)A + \gamma L |A|^2A =[\delta-i\alpha]A+i\theta E_{in},\label{LLE}\\
& d(i \partial_t)=\sum_{n\geq 1}\frac{\beta_n}{n!}(i\partial_t)^n,\\
&D(\omega)=\sum_{n \geq 1}\frac{\beta_n}{n!}\omega^n;
\end{align}
where $z$ is the longitudinal coordinate, $t$ is the time in a reference frame moving at the  velocity of the pump ($t\rightarrow t-\frac{z} {V}$), $A$ is the field envelope, $E_{in}$ is the input field, $L$ is the cavity length, $\theta$ is the transmissivity of the mirror, $\delta$ is the cavity detuning, $\gamma$ is the nonlinear coefficient, $\alpha$ accounts for the total cavity losses, directly linked to the finesse $\mathcal{F}$: $\alpha=\pi/\mathcal{F}$ (valid for $\mathcal{F}\gg1$), $\beta_n~(n \geq 2)$ account for the group velocity dispersion (GVD for n=2) and HOD. Here, $\beta_1$ takes into account the synchronization mismatch that might occur between the pulsed pump and the cavity repetition rate \cite{coen_convection_1999,Negrini2023_synchronization}. Hence, $\beta_1$ is defined as $\beta_1 = -\frac{\Delta T}{L} - \frac{1}{V}$, where $\Delta T$ stands for the difference between the actual cavity round-trip time and the ideal cavity round-trip time. In the pulsed pumping scheme case with a repetition rate $f_{rep}$ and a cavity with a free spectral range $FSR$ \cite{coen_convection_1999,Negrini2023_synchronization}, we have $\Delta T = \frac{1}{f_{rep}} - \frac{1}{FSR}$.
We search for stationary solutions of Eq.~(\ref{LLE}), i.e. we impose $\partial_z=0$ :
\begin{equation}
 L  d(i\partial_t)A + \gamma L |A|^2A =[\delta-i\alpha]A+i\theta E_{in}.\label{steady_LLE}\\
\end{equation}
The  continuous wave (CW) or homogeneous steady states (HSS) $A_s=A(z,t)$ are the solution of the cubic equation
\begin{equation}
\theta^2P_{in}=P\left((\gamma L -\delta)^2+\alpha^2\right),
\end{equation}
where $P=|A_s|^2$.
In order to characterize how two different HSSs are connected by a front, we perturb a HSS and analyse the behavior of the perturbation. We insert $A(t)=A_s+\eta(t)$ in Eq.~(\ref{steady_LLE}) and linearise assuming $\eta$ small ($|A_s| \gg \eta$). We obtain:
\begin{equation}\label{lin}
L d(i\partial_t)\eta + (2 \gamma L |A_s|^2 -\delta+i\alpha) \eta + \gamma L  A_s^2 \eta^*  =0.
\end{equation}
The solution of Eq.~(\ref{lin}) is searched in the form $\eta(t)~=~a~e^{\lambda t}~+~b^*e^{\lambda^* t}$, with $a$, $b$, $\lambda=\sigma+i\omega$ complex constants. By collecting the exponentials having the same power, we get the system: 
\begin{align}\label{system}
\nonumber  & M (a,b)^T=0, \\
& \scriptsize
M=\left[
\begin{array}{ c c}
 D(i \lambda) L+2 \gamma L |A_s|^2 -\delta+i\alpha &  \gamma L  A_s^2\\
 \gamma L  A_s^{*2} & D(-i \lambda) L +2 \gamma L |A_s|^2 -\delta-i\alpha)
\end{array} \right]
\end{align}
In order to get a nontrivial solution, we impose $\det(M)=0$, which gives:
\begin{equation}\label{disp}
\small
\left(D_e(i\lambda)L+2\gamma P L-\delta\right)^2-\left(i\alpha+D_o(i\lambda)L\right)^2-(\gamma P L)^2=0,
\end{equation}
where we have defined the even and the odd parts of the complex dispersion relation as:
\begin{align}
D_e(i\lambda)=\frac{D(i\lambda)+D(-i\lambda)}{2},\\
D_o(i\lambda)=\frac{D(i\lambda)-D(-i\lambda)}{2}.
\end{align}
%
%
Fig.~\ref{fig:SWtheo}(e) gives a representation of Eq.~(\ref{disp}). It showcases the real part (i.e. $\operatorname{Re}(\det(M))=0$) in green and the imaginary part (i.e. $\operatorname{Im}(\det(M))=0$) in orange. The roots, either real or in complex-conjugate pairs ($\lambda=\sigma+i\omega$), correspond to the intersection of these curves (blue dots). The real part ($\sigma$) characterizes the exponentially growing or decaying behavior in the transverse dimension (i.e. in time), while the imaginary part ($\omega$) denotes the oscillatory nature of the perturbations around the continuous wave solution.
Therefore, the transverse linear stability analysis reveals that the oscillatory tails are an inherent characteristic of the continuous wave (CW) solution itself, and the frequency of these oscillations is determined by the imaginary part of the roots of Eq.~(\ref{disp}). Interestingly, we found that Eq.~(\ref{disp}) can be written as
\begin{equation}\label{disp_pm}
\small
i\alpha+D_o(i\lambda)L=\pm\sqrt{\left(D_e(i\lambda)L+2\gamma P L-\delta\right)^2-(\gamma P L)^2}\;,
\end{equation}
In the limit of $|\gamma P L| \ll |D_e(i\lambda)L+2\gamma P L-\delta|$, we can expand Eq.~(\ref{disp_pm}) to get the following equation
\begin{equation}\label{taylor_disp_pm}
i\alpha+D_o(i\lambda)L\pm[\left(D_e(i\lambda)L+2\gamma P L-\delta\right]=0.
\end{equation}
The positive branch of Eq.~(\ref{taylor_disp_pm}) can be written in the form of a phase-matching condition
\begin{equation}\label{phase_matching}
L\sum_{n\ge1}\frac{\beta_n}{n!}(i\lambda)^n -i\lambda\frac{L}{V}+2\gamma P L-\delta+i\alpha=0,
\end{equation}
which has been used before to describe emission of radiation from solitons (bright and dark) or shock waves~\cite{Malaguti:14}. 
Even if the equation is the same, it has been derived assuming that a nonlinear localised solution emits DWs due to perturbations (i.e. HOD). 
The derivation of Refs.~\cite{Malaguti:14,macnaughtan_temporal_2023,xu_frequency_2021,ji_engineered_2023,xiao_modeling_2023,li_experimental_2023} do not justify the use of Eq.~(\ref{phase_matching}) to  characterise the nonlinear solution itself, namely its oscillatory tails which give rise to spectral shoulders. 
The procedure reported here does not make use of this assumption, giving clear and solid physical foundations to the derivation of Eq.~(\ref{phase_matching}). 

Fig.~\ref{fig:SWtheo} illustrates an example of the analytical predictions. The parameters are L$=16.18$~cm, $\mathcal{F}=600$ $\gamma=10.8~W^{-1}km^{-1}$, $\delta=0.032$~rad, $P_{in}=0.5$~W, which correspond to experimental ones used in Part.~\ref{sw}. A down SW is represented in Fig.~\ref{fig:SWtheo}(a). It connects the higher and lower states of the bistable regime of the system [Fig.~\ref{fig:SWtheo}(b)], and is therefore limited to this regime [green arrow in Fig.~\ref{fig:SWtheo}]. As expected the oscillation frequency of the SW corresponds to the imaginary part of the roots of the polynomial~(\ref{disp}) represented in Fig.~\ref{fig:SWtheo}(e). Note that as the oscillations appear on the lower state, the power of the lower branch of the bistable regime $P_L$ has thus to be considered in Eq.~(\ref{disp}). In this example, the stationary states calculation give $P_L=0.75$~W. By incorporating it into Eq.~(\ref{disp}), we find $f_{SW}=2.67$~THz. 

SWs propagate indefinitely in  a resonator only if their velocity relative to the driving field is zero \cite{coen_convection_1999,parra-rivas_origin_2016,parra-rivas_coexistence_2017,anderson_zero_2022}. Otherwise, they eventually decay on the lower or upper HSS.
For each value of the detuning, this condition is achieved for only one  value of the input power $P_{in}$, corresponding to a single couple of stable higher and lower state values ($P_H$ and $P_L$). This point, named Maxwell point (MP) \cite{parra-rivas_origin_2016,parra-rivas_coexistence_2017,anderson_zero_2022}, can be found by calculating the SW velocity through numerical methods \cite{parra-rivas_coexistence_2017} [Fig.~\ref{fig:SWtheo}(c), (d)]. In the example shown in Fig.~\ref{fig:SWtheo}, the SW [Fig.~\ref{fig:SWtheo}(a)] is at its MP and therefore has zero speed [orange dotted line in Fig.~\ref{fig:SWtheo}(a), (b), (c) and (d)]. 

In the following, Eq.~(\ref{disp}) is used to predict OFC generation via SWs excitation in FFP resonators. Note that the cross-phase modulation (XPM) which occurs in these cavities between the co and counter propagating fields is considered as an additional phase shift and can be compensated for by a slight change in the pump laser detuning: $\delta \approx \delta_{FP}-4f\gamma PL$ \cite{ziani_theory_2023,bunel_impact_2023} (where $f$ is the ratio between the pulse duration and the cavity roundtrip time). For the experimental results presented in the following, this phase difference is very small and always around $0.003$~rad.

\section{\label{setup}Experimental setup}

\begin{figure}
\includegraphics{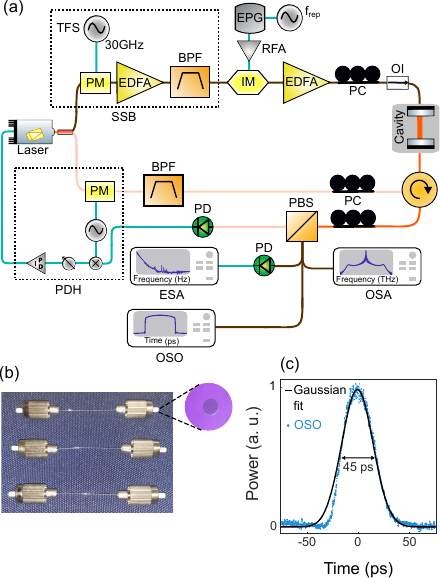}
\caption{\label{fig:setup}Experimental setup and FFP cavities. (a) Experimental setup with a two-arms stabilization system. Brown line: nonlinear beam; beige line: control beam. Both beams are perpendicularly polarized to each other. TFS: Tunable Frequency Synthesizer; EPG: Electrical Pulse Generator; RFA: Radio Frequencies Amplifier; IM: Intensity Modulator; PM: Phase Modulator; EDFA: Erbium Doped Fiber Amplifier; PC: Polarization Controller; OI: Optical Isolator; PD: Photodiode; PBS: Polarization Beam Splitter; PDH: Pound-Drever-Hall; SSB: single-side-band generator; ESA: Electrical Spectrum Analyser; OSA: Optical Spectrum Analyser; OSO: Optical Sampling Oscilloscope. (b) Photograph of FFP cavities with zoom on a deposited mirror. (c) Recording of a pump pulse with an OSO (blue dotted line) and the the fitted Gaussian curve (black line)}
\end{figure}

\begin{figure*}
\includegraphics{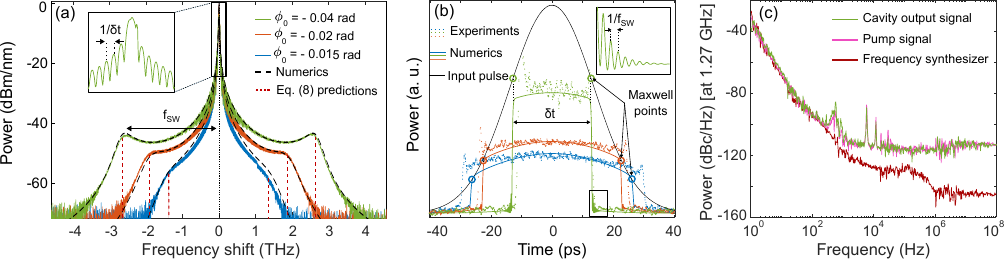}
\caption{\label{fig:SW}Experimental recordings performed in cavity $\#2$ as a function of the cavity detuning. Blue: $\delta=0.012$~rad; orange: $\delta=0.017$~rad; green: $\delta=0.032$~rad. (a) Generated spectra recorded with an OSA. Colored lines: experiments; black dashed lines: numerics. (b) Time domain traces recorded with an OSO. Black line: Pump pulse envelope; colored dotted lines: time-domain recording corresponding to the experimental spectra in (a); colored lines: corresponding numerics; colored circles: Maxwell points (c) Phase noise spectrum measurements. Red: Frequency synthesizer signal; pink: pump pulse train; green: output signal when $\delta=0.035$~rad (green trace in (a) and (b)).}
\end{figure*}

\begin{table}
\caption{\label{tab:tablCavity}Cavities parameters}
\begin{ruledtabular}
\begin{tabular}{cccc}
&Cavity $\#1$&Cavity $\#2$&Cavity $\#3$\\
\hline
L (cm)&8.21&8.09&8.78\\
FSR (GHz)&1.28&1.27&1.17\\
$\mathcal{F}$&500&600&420\\
$\gamma$ (W$^{-1}$km$^{-1}$)&2&10.8&10.8\\
$\beta_2$ (ps$^2$km$^{-1}$)&6.3&1.3&0.4
\end{tabular}
\end{ruledtabular}
\end{table}

Three FFP cavities with  similar lengths (around $8$~cm) were fabricated from normal dispersion optical fibers with different GVD values: $6.3$~ps$^2.$km$^{-1}$, $1.3$ ps$^2.$km$^{-1}$ and $0.4$~ps$^2.$km$^{-1}$, and are represented in Fig.~\ref{fig:setup}(b). For all cavities, we checked numerically that the respective fiber HOD terms does not have a significant impact. They are made by connecting the optical fibers between two FC/PC connectors, where Bragg mirrors are deposited at each sides [zoom in Fig.~\ref{fig:setup}(b)] with a physical vapor deposition technique, to achieve $99.86\%$ reflectance over a $100$~nm bandwidth at the pump wavelength ($1550$~nm) \cite{zideluns_automated_2021}. Hence, they possess a free spectral range (FSR) around $1.2$~GHz, and a finesse ranging between 400 and 600 (i.e. a quality-factor between $69$ and $92$~million). For clarity and consistency throughout the text, they are referred to as cavity $\#1$, $\#2$ and $\#3$ and their parameters are listed in Table~\ref{tab:tablCavity}. As reported, these devices are very practical for generating OFCs in the GHz range \cite{li_experimental_2023,li_ultrashort_2023,jia_photonic_2020,apl_soliton,bunel_observation_2023}. 
Nonetheless, due to their few cm length, these systems are sensitive to mechanical and acoustical vibration, and thermal fluctuations, requiring a stabilization or isolation. For this study, we employed an active laser stabilization system to counteract cavity resonance variations, a technique that has shown excellent results in previous studies \cite{apl_soliton,bunel_impact_2023,englebert_temporal_2021}. The FFP resonators are exploited in the experimental setup described in Fig.~\ref{fig:setup}(a). Similar to studies on fiber ring cavities, that stabilize the pump laser in one direction while inducing nonlinear effects in the other by using clockwise and counter-clockwise propagation directions \cite{leo_temporal_2010,Negrini2023_synchronization,englebert_temporal_2021}, a two-arms stabilization scheme is utilized. However, the two main polarization axes provided by the natural birefringence of the fiber cavities are used, rather than managing the two-way flow which is impractical due to the Fabry-Perot structure. In Fig.~\ref{fig:setup}(a), the lower beige arm employs a Pound-Drever-Hall system \cite{black_introduction_2001,drever_laser_1983} to lock the pump laser on the top of a cavity resonance, while the upper brown arm triggers nonlinear effects by pumping the cavity with a Gaussian pulse train. This pulse train is generating using an intensity modulator driven by an electrical pulse generator to obtain Gaussian pulses with a duration of $45$~ps at full width half maximum [Fig.~\ref{fig:setup}(c)]. A frequency synthesizer enables precise adjustment of the repetition rate [$f_{rep}$ in Fig.~\ref{fig:setup}(a)], allowing it to match the cavity roundtrip time within the Hz range. Finally, a homemade tunable single-sideband generator [see SSB in Fig.~\ref{fig:setup}(a)] affords fine control over the cavity detuning within a range of hundreds of Hz, which is three order of magnitude lower than the linear transfer function linewidth.

\section{\label{sw}Switching waves generation}

Our setup facilitates the straightforward generation of SWs, by making possible pulsed pumping and perfect control of the cavity detuning. When the latter is such that the system is bistable, SWs can be generated at the MP. Additionally, employing a pulsed-pump with sufficiently high peak power ensures a specific time on the rising and falling fronts of the pump corresponding to the MP, where SWs can effectively link the lower and higher states of the cavity without any group-velocity offset, and thus remain stable \cite{anderson_zero_2022}. Fig.~\ref{fig:SW}(a) and (b) report cavity output signals in the frequency and time domains, respectively, with Cavity $\#2$, for different detuning values. The generation of SWs is clear from its signature in the spectral domain, i.e. symmetrical shoulders around the pump [Fig.~\ref{fig:SW}(a)] corresponding to the SWs frequency and leading the OFC span. In the time domain, the SWs generation is characterized by the steep pulse fronts [Fig.~\ref{fig:SW}(b)], which link the higher and lower states. It results in almost-square output pulses, which can be seen in the spectra with a sinus cardinal like shape modulation [see insert in Fig.~\ref{fig:SW}(a)]. These observations illustrate also that the SWs frequency, and thus the generated OFC width, increases with the detuning [Fig.~\ref{fig:SW}(a)], from $1.3$~THz for $\delta=0.012$~rad, to $2.7$~THz for $\delta=0.032$~rad. It is also noticeable that the spectral shoulders become more marked with larger detuning values, leading to increased amplitude in the oscillatory tails of SWs. 
Interestingly, the MP rises the top of the driving pulse with the detuning [circles in Fig.~\ref{fig:SW}(b)], leading to much shorter pulse duration at the cavity output, in accordance with other studies in ring cavities \cite{anderson_zero_2022,coen_convection_1999}. Pulses can be obtained with a duration from $52$~ps to $25$~ps. All these observations are in very good agreement with numerical of LLE [solid lines in Fig.~\ref{fig:SW}(b)]. The SWs frequencies and the shoulders positions can also be analytically predicted thanks to the model introduced in Part~\ref{theory}. The red dashed lines in Fig.~\ref{fig:SW}(a) correspond to the imaginary part of the roots of the Eq.~(\ref{disp}). They perfectly match with the shoulders positions. As an example, Fig.~\ref{fig:SWtheo} depicts the case when the detuning is set to $0.032$~rad [green plots in Fig.~\ref{fig:SW}]. The MP correspond to $P_L=0.75$~W, and leads to $f_{SW}=2.67$~THz (roots of Eq.~(\ref{disp})). 
Note that this oscillations are not observed experimentally with the OSO due to its limited bandpass (700~GHz). They can be seen through numerics [insert in Fig.~\ref{fig:SW}(b)]. 

Beyond enabling the generation of OFCs through spectral broadening, SWs are particularly interesting due to their excellent stability, which make them very attractive for applications. Phase noise spectra measurements revealed that the nonlinear process do not introduce any noise [green and pink lines in Fig. \ref{fig:SW}(c)]. A phase noise under $-110$~dBc/Hz from $500$~Hz is observed for the generated OFC signal [green line in Fig. \ref{fig:SW}(c)]. When compared to the reference frequency synthesizer [red line in Fig. \ref{fig:SW}(c)], there is no additional phase noise observed in the lower frequencies, indicating an excellent transmission of stability during the comb generation process. Only in the higher frequencies, a phase noise difference of $30$~dB is noted [green and red lines in Fig. \ref{fig:SW}(c)].

\section{\label{mismatch}Impact of the synchronisation mismatch}

\begin{figure*}
\includegraphics{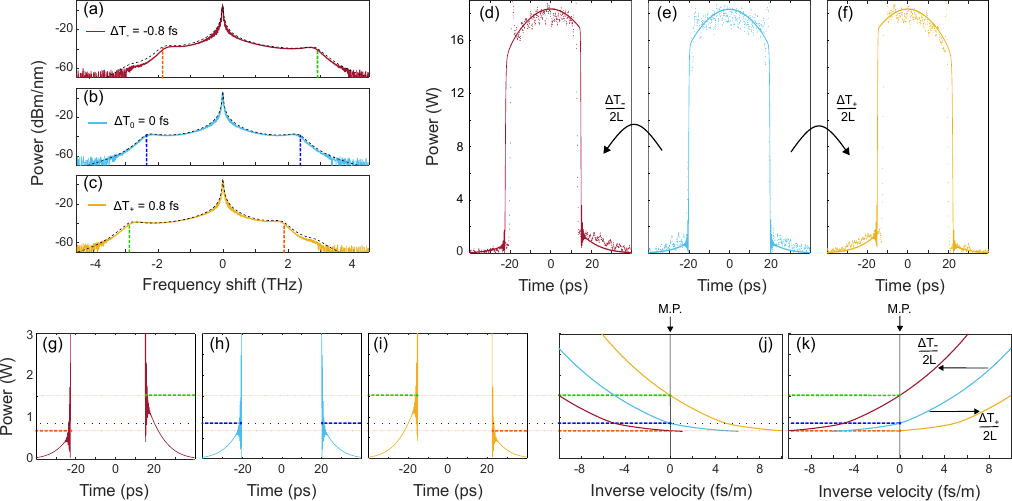}
\caption{\label{fig:beta1}Switching waves with synchronization mismatch. Color code for all figures: yellow: $\Delta T=0.8$~fs, blue: $\Delta T=0$~fs, red: $\Delta T=-0.8$~fs. (a), (b) and (c) colored lines: experimental spectra for the different synchronization mismatch values; black dashed lines: corresponding numerics from LLE. (d), (e) and (f) Corresponding time domain traces. Dots: Experimental traces recorded with an OSO; lines: Numerics from LLE. (g), (h), (i) Zoomed-in views are provided for better visibility of the oscillation power levels at the pulse bottoms. (j) and (k) SW inverse velocity as a function of the lower state power, for the three synchronization mismatch values; (j): rising front, (k): falling front. Colored dashed lines are used to identify which spectral component corresponds to which front.}
\end{figure*}

The synchronization mismatch between the pump repetition rate and the cavity roundtrip time significantly influences the dynamics of SWs \cite{coen_convection_1999,xu_frequency_2021,macnaughtan_temporal_2023,xiao_modeling_2023}. It is defined as the difference between the pump period and the cavity roundtrip time: $\Delta T = \frac{1}{f_{rep}} - \frac{1}{FSR}$.
Here, spectra were obtained for three distinct cavity mismatch values ($-0.8$, $0$, and $0.8$ fs) with Cavity $\#2$,
while maintaining a fixed detuning $\delta=0.025$~rad. Without synchronisation mismatch ($\Delta T=0$), the spectrum and output pulses are perfectly symmetric. However, with positive (negative) $\Delta T$, the spectrum's shoulders shift towards higher (lower) frequencies, resulting in asymmetrical spectra. This is experimentally observed in Fig.~\ref{fig:beta1}(a), (b) and (c) [colored lines] in good agreement with numerics [black dashed lines], where synchronisation mismatch is considered in the governing equation LLE [Eq.(\ref{LLE})] by introducing a drift velocity \cite{coen_convection_1999,Negrini2023_synchronization} $\frac{1}{\beta_1} =\left( -\frac{\Delta T}{2L} - \frac{1}{V}\right)^{-1}$.  Corresponding time domain traces, recorded with an OSO, and calculated from numerics are shown in Fig.~\ref{fig:beta1}(d), (e), and (f). They also exhibit asymmetry due to the synchronisation mismatch ($\Delta T \ne 0$). Notably, the frequency of oscillations tails of the fronts, reflecting the spectrum's shoulder positions, differs along with the power at which these oscillations appear [Fig.~\ref{fig:beta1}(g), (h), and (i)]. The model depicted in Part~\ref{theory} gives a clear explanation of this phenomenon. First, the power at which the oscillations appear can be determined finding the MPs for each front. This is achieved by calculating the SWs inverse velocity as a function of the lower state power $P_L$ for both the rising [Fig.~\ref{fig:beta1}(j)] and falling [Fig.~\ref{fig:beta1}(k)] fronts, considering the three $\Delta T$ values. An inverse velocity contribution of $\Delta T/(2L)$ is added when $\Delta T \ne 0$. As depicted, the yellow curves in Fig.~\ref{fig:beta1}(j) and (k) are shifted by $\Delta T_{+}/(2L)=5$~fs/m compared to the blue curves, while the red curves is shifted by $\Delta T_{-}/(2L)=-5$~fs/m. As a result, the power levels where the SWs velocity reaches zero (i.e., the MP) vary with the synchronization mismatch. Additionally, given that the slopes of the curves are inverse for the rising and falling fronts, the MP for the two fronts differ when $\Delta T \ne 0$. For instance, for $\Delta T_{+} = 0.8$~fs [yellow case in Fig.~\ref{fig:beta1} (c), (f) and (i)], the MP corresponds to $P_L=1.55$~W for the rising edge [green dotted lines in Fig.~\ref{fig:beta1} (i) and (j)], while it is $P_L=0.67$~W for the falling edge [orange dotted lines in Fig.~\ref{fig:beta1} (i) and (k)]. Consequently, the $P_L$ value used to compute the SWs frequency in Eq.~(\ref{disp}) are different, explaining the difference in SW frequencies for both fronts and the asymmetry in the spectral domain. Specifically, we found $-2.88$~THz for the rising edge and $1.81$~THz for the falling one [orange and green dotted lines in Fig.~\ref{fig:beta1} (c)]. There is a very good agreement between both numerical simulations and experimental observations. When $\Delta T = 0$ [blue case in Fig.~\ref{fig:beta1} (b), (e) and (h)], the MP remains the same for both fronts, allowing us to calculate the SWs frequency as detailed in Part~\ref{sw}. Conversely, for $\Delta T_{-} = -0.8$~fs (red case in Fig.~\ref{fig:beta1} (a), (d) and (g)) a perfectly opposite scenario to that of $\Delta T_{+} = 0.8$~fs occurs. Here, the two fronts are reversed, with the MP registering at $0.67$~W and $1.55$~W for the rising and falling edges, respectively, corresponding to shoulder positions at $-1.81$~THz and $2.88$~THz.

\section{\label{disper}Impact of the group velocity dispersion}

\begin{figure}
\includegraphics{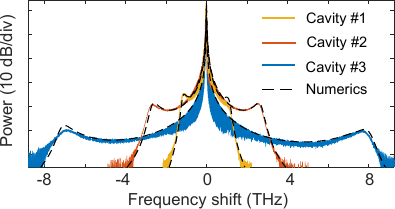}
\caption{\label{fig:beta2}OFC generation as a function of the cavity dispersion. Cavity~$\#$1: $\delta=0.032$~rad, $P_{in}=3$~W. Cavity~$\#$2: $\delta=0.032$~rad, $P_{in}=0.5$~W. Cavity~$\#$3: $\delta=0.09$~rad, $P_{in}=1.5$~W.}
\end{figure}

Eq.~(\ref{disp}) and (\ref{phase_matching}) predicts that the cavity dispersion ($\beta_2$) also impacts the SWs frequencies and thus the OFC span. To illustrate it, OFCs were similarly generated in two other cavities (Cavity $\#1$ and $\#3$), exhibiting different GVD values. The results are depicted in Fig.~\ref{fig:beta2}. In this comparison, the analysis is primarily qualitative, considering the finesse and nonlinear coefficients across different cavities. The pumping power and detuning value are adjusted to achieve comparable $\phi$ values, where $\phi=2\gamma P(2L) -\delta$, ensuring that only the dispersive term changes in Eq.(\ref{disp}) and (\ref{phase_matching}). The setup parameters employed for generating the respective OFCs are detailed in the Figure~\ref{fig:beta2}'s caption. Thus, $\phi_{\#1}=0.004$, $\phi_{\#2}=0.006$, and $\phi_{\#3}=0.004$, making them comparable in relation to the dispersion term values $\frac{\beta_2}{2}\omega^2$, more than 500 times greater at $4$~THz for Cavity $\#2$ for instance. Note that dispersion exerts the most significant influence, often serving as the sole determinant in numerous studies for predicting the spectral positioning of dispersive phenomena \cite{brasch_photonic_2016,anderson_zero_2022,kippenberg_dissipative_2018,moille_ultra-broadband_2021}. Notably, the cavity with the lowest dispersion showcases a remarkable capability to generate a wide-ranging OFC spanning a $15$~THz bandwidth. In this instance, the flattened dispersion curve of the fiber used to construct Cavity~$\#$3 leads to a weak contribution of third order of dispersion (TOD) which would leads to an asymmetry in the spectrum. Nonetheless, TOD cannot be completely neglected, leading to a very low $\beta_3$ value ($\beta_3=-2.73 \times 10^{-3}$~ps$^3$km$^{-1}$) and a slight asymmetry in the  spectrum, $f_{SW}=-7$~THz for the left front VS $7.9$~THz for the right front [blue line in Fig.~\ref{fig:beta2}]. We observe that radiation efficiency decreases with SW frequency, given that the cavity with the highest dispersion get the most powerful comb lines. 
These findings further underscore the inherent benefits of FFP cavities, highlighting their manufacturing adaptability to meet specific needs.

\section{Discussion and conclusion}
We demonstrate experimentally that broadband OFCs can be generated in fiber Fabry-Perot resonators in the normal dispersion regime, by generating SWs from a pulsed pump. We illustrate the impact of the dispersion using different cavities as well as the impact of the synchronisation mismatch between the pump and the cavity. We introduce a theoretical approach based on the transverse linear stability analysis, which allows us to predict the bandwidth of the OFC, \textit{i.e. }the oscillation frequency of the SWs. Unlike previous studies that attribute SW generation to dispersive wave emission from an impulse front \cite{macnaughtan_temporal_2023,xu_frequency_2021,ji_engineered_2023,xiao_modeling_2023,li_experimental_2023}, our work considers that SWs are exact solutions of the Lugiato-Lefever equation. Remarkably, we have derived a phase-matching condition similar to those employed in prior studies, enabling us to predict the SW frequency with high accuracy. In addition to providing insights into the dynamics of SW formation, our results highlight the ability of FFP resonators to produce broad and stable frequency combs in the GHz range. By fine-tuning the dispersion and nonlinearity characteristics of the optical fiber used, we have realized optical frequency combs that cover up to 15~THz in bandwidth, feature a frequency spacing of 1.2~GHz, and demonstrate phase noise as low as -110~dBc/Hz at 1~kHz. This research contribute to a better understanding of the dynamics of SWs in high quality-factor resonators and facilitates the development of innovative platforms for OFC generation.

\begin{acknowledgments}
The present research was supported by the agence Nationale de la Recherche (Programme Investissements d’Avenir, ANR FARCO, ANR TRIPLE, LABEX CEMPI); Ministry of Higher Education and Research; European Regional Development Fund (WAVETECH),  the CNRS (IRP LAFONI); Hauts de France Council (GPEG project), and the university of Lille (LAI HOLISTIC)
\end{acknowledgments}

\section*{Data Availability Statement}

The data that support the findings of this study are available from the corresponding author upon reasonable request.

\nocite{*}

\bibliography{references}

\end{document}